# Active galactic nuclei as scaled-up Galactic black holes

I. M. M^cHardy[1], E. Koerding[1], C. Knigge[1], P. Uttley[2] & R. P. Fender[1]

[1]School of Physics and Astronomy, The University, Southampton SO17 1BJ, UK.

[2]Astronomical Institute Anton Pannekoek, University of Amsterdam, Kruislaan 403, 1098 SJ Amsterdam, The Netherlands.

**A long-standing question is whether active galactic nuclei (AGN) vary like Galactic black hole systems when appropriately scaled up by mass[1-3]. If so, we can then determine how AGN should behave on cosmological timescales by studying the brighter and much faster varying Galactic systems. As X-ray emission is produced very close to the black holes, it provides one of the best diagnostics of their behaviour. A characteristic timescale, which potentially could tell us about the mass of the black hole, is found in the X-ray variations from both AGN and Galactic black holes[1-6], but whether it is physically meaningful to compare the two has been questioned[7]. Here we report that, after correcting for variations in the accretion rate, the timescales can be physically linked, revealing that the accretion process is exactly the same for small and large black holes. Strong support for this linkage comes, perhaps surprisingly, from the permitted optical emission lines in AGN whose widths (in both broad-line AGN and narrow-emission-line Seyfert 1 galaxies) correlate strongly with the characteristic X-ray timescale, exactly as expected from the AGN black hole masses and accretion rates. So AGN really are just scaled-up Galactic black holes.**

The first detailed observations of AGN X-ray variability showed scale-invariant behaviour on all timescales from approximately days to minutes[8,9], with no characteristic timescale from which black hole masses ($M_{BH}$) might be deduced. However, subsequent observations[1-3] showed that, on longer timescales, a characteristic timescale could be derived from the power spectral densities (PSDs; that is, variability power, $P(\nu)$, at frequency, $\nu$, or timescale, $1/\nu$) of the X-ray light curves.

All AGN PSDs are best fitted on long timescales by a powerlaw of slope −1 ($P(\nu) \propto \nu^{-\alpha}$ with $\alpha \approx 1$) which breaks to a steeper slope ($\alpha > 2$) on timescales shorter than a 'break' timescale, $T_B$. For some AGN, the $\alpha \approx 1$ slope can be followed to long timescales for >3 decades with no further break, similar to Galactic black hole X-ray binary systems (GBHs) in their 'soft' states[5,7,10-13]. For other AGN, the slope can only be followed for <2 decades, which is insufficient to distinguish them from GBHs in their 'hard' states where, in the power-law description of the PSD, a second break, to slope $\alpha \approx 0$, is seen ~1.5–2 decades below the $\alpha \approx 1$–2 break. Here we use the timescale associated with the $\alpha \approx 1$–2 break as a characteristic timescale, irrespective of likely state. The reason for the sudden decrease in variability power on timescales shorter than $T_B$ is not clear, but the variability probably originates within the accretion disk[14] surrounding the black hole and $T_B$ may be associated with the inner edge of the disk.

A major difficulty in establishing a quantitative timing link between AGN and GBHs has been the large scatter in the $M_{BH}$–$T_B$ relationship[7]. In particular, for a given $M_{BH}$, the high accretion rate narrow line Seyfert 1 galaxies (NLS1s) have smaller values of $T_B$ than other AGN[5,11]. We therefore suggested that $T_B$ is inversely dependent on a second variable, possibly the black hole spin, but probably the accretion rate[5,11,12] (often written $\dot{m}_E$, in units of the maximum possible, Eddington, accretion rate). Other researchers, estimating $T_B$ from the excess variance in the X-ray light curves, also find a correlation of $T_B$ with $M_{BH}$ but the dependence on $\dot{m}_E$ is not clear[6,13]. Here, we properly quantify the relationship between $T_B$, $M_{BH}$ and $\dot{m}_E$. To make the parameters properly independent, we fit to the observable quantitity, the bolometric luminosity, $L_{bol}$, rather than $\dot{m}_E$, although for the objects discussed here, which are radio-quiet, $\dot{m}_E$ is well approximated by $L_{bol}/L_E$ (Eddington luminosity

$L_E \propto M_{BH}$). Motivated by the rough linear scaling between $T_B$ and $M_{BH}$, and the rough decrease of $T_B$ with increasing $\dot{m}_E$, we hypothesize that:

$$\text{Log } T_B = A \log M_{BH} - B \log L_{bol} + C$$

and determine the best-fit values of $A$, $B$ and $C$ from a simple parameter grid search. (Details of PSD parameterization, fitting procedure and date relevant to this paper are given in the Supplementary Information.) We first fit to the 10 AGN that have well measured PSD breaks and reasonable measurements of mass and bolometric luminosity[12]. This sample contains a range of accretion rates from $\dot{m}_E > \sim 1$ (Akn 564) to $\sim 10^{-3}$ (NGC 4395). The fit (Fig. 1) is good with $A = 2.17^{+0.32}_{-0.25}$, $B = 0.90^{+0.3}_{-0.2}$ and $C = -2.42^{+0.22}_{-0.25}$.

To determine whether the same scaling relationship extends to GBHs, we include two bright GBHs (Cyg X-1 and GRS 1915+105) in radio-quiet states where, for proper comparison, their high frequency PSDs are well described by the same cut-off ('breaking') power-law model which best describes AGN[10,15], and where broad band X-ray flux provides a good measurement of bolometric luminosity. For Cyg X-1, we combined measurements[10] of $T_B$ with simultaneous measurement of the bolometric luminosity[16] over a range of luminosities. For GRS 1915+105, we measured an average $T_B$ from the original X-ray data, and determined $L_{bol}$ from the published fluxes[15] and generally accepted distance (11 kpc).

Using Cyg X-1, we can test whether the AGN scaling relationship applies within any one object (that is, at fixed mass). We bin into 5 luminosity bins and find that $T_B \propto L_{bol}^{-1.3\pm0.2}$. This dependence is consistent with that for AGN, although slightly larger, probably because the bolometric range covered includes the transition zone between hard and soft states where more rapid changes of timescale with luminosity occur than in either state alone[17].

It is particularly important to determine whether these Galactic data can be fitted, self-consistently, with the AGN data. We first fit to Cyg X-1 and the AGN and find excellent agreement with the AGN alone. Including also GRS 1915+105, whose break timescale is ~10× shorter than for Cyg X-1, we again find perfect agreement (Fig. 1) with $A = 2.10 \pm 0.15$, $B = 0.98 \pm 0.15$ and $C = -2.32 \pm 0.2$. As the fit is good, no unknown source of error need be invoked, implying that no other parameter (such as spin) has as large an effect on $T_B$ as $M_{BH}$ or $\dot{m}_E$. Thus we answer a long-standing question and show—using a self-consistently derived set of AGN and GBH timing data—that, over a range of ~$10^8$ in mass and ~$10^3$ in accretion rate, AGN behave just like scaled-up GBHs. Assuming $\dot{m}_E \approx L_{bol}/L_E$, then $T_B \approx M_{BH}^{1.12}/\dot{m}_E^{0.98}$.

If the break timescale is proportional to a thermal or viscous timescale associated with the inner radius of the accretion disk, $R_{disk}$, then from our fit we expect $R_{disk} \propto \dot{m}_E^{-2/3}$. Models based on evaporation of the inner disk[18] predict $R_{disk} \propto \dot{m}_E^{-0.85}$ and $T_B \propto M_{BH}^{1.2}$, which is not too far from our observations.

In their hard states, GBH PSDs are often better described by the combination of two or more lorentzian-shaped components[19] whose timescales approximate the break timescales in the power-law description of the PSD. Although not included in our fits as a number of assumptions are required, it has already been shown, from combined radio and X-ray observations[20], that in the hard state of GBH GX339–4 the timescale associated with the lorentzian closest to the $\alpha \approx$ 1–2 break timescale varies as ~$1/\dot{m}_E$ (see Supplementary Information for more details). Combined radio and X-ray luminosities also underpin another, non-timing, analysis, from which a strong scaling, particularly of jet properties, between AGN and GBHs has recently been shown[21,22].

In Fig. 2 we show a projection of the $T_B$–$M_{BH}$–$\dot{m}_E$ plane and all objects lie close to it. Thus the same process (for example, accretion rate variations[14]) probably produces X-ray variations in the same way in all accreting black holes. Thus, although we agree completely with previous criticism[7] that $T_B$, on its own, is not a precise indicator of $M_{BH}$, we show that $T_B$, when combined with an estimate of $\dot{m}_E$, is a very good mass indicator. This indicator may have particular value where mass determination is otherwise difficult—for obscured AGN and for potential intermediate mass ($10^3$–$10^4$ solar masses) black holes, for example.

Our results establish that $T_B$ is a powerful tracer of the innermost accretion processes in AGN and GBHs. These processes produce photons that must affect the larger scale AGN properties, but no relationship has yet been found between $T_B$ and such properties. One particularly important property, often used to classify AGN, is the width of the permitted optical emission lines. These lines are narrower in more X-ray variable AGN[23,24], but the reason has never been satisfactorily explained.

Theoretically, a correlation between $M_{BH}$, $\dot{m}_E$ and linewidth is expected from simple scaling relationships for the gas surrounding the black hole (the broad line region, BLR) from which emission lines, broadened by Doppler velocities, $V$, originate. We assume that the emission lines result from photoionization and that the ionizing luminosity $L \propto M_{BH}\dot{m}_E$. We also assume virial motion for the BLR gas: that is, $GM_{BH}/R_{BLR} \approx V^2$, where $R_{BLR}$ is the inner radius of the BLR, and $R_{BLR} \propto L^a$. For the 'locally optimized condition' for production of emission lines by gas at the same optimum density and ionization state, we expect $a = 0.5$, and the most recent observational study[25] finds $a = 0.518 \pm 0.039$. If $a = 0.5$, we expect $V^4 \approx M_{BH}/\dot{m}_E$.

A strong independent test of our derivation that $T_B \propto M_{BH}/\dot{m}_E$ is therefore provided by plotting (Fig. 3) $T_B$ versus $V$, for all 9 of our primary sample of 10 AGN with well measured PSD breaks, for which measurements of the FWHM of the variable, broad component, of the Hβ line, $V$, exist[26]. We note a much tighter relationship than in any previous correlations between linewidth and other parameters quantifying variability—for example, fractional variability of the optical continuum[27] or r.m.s. X-ray variability[23,24]. Both $T_B$ and $V$ are observables, so this relationship depends on no assumptions.

Parameterizing $\log T_B = D\log V + E$, we perform a grid search for the best-fit values of $D$ and $E$. Measurement errors on $V$ are typically ~10–15%, but differences of >50% can occur between different epochs[26]. Thus large time differences between measurements of $T_B$ and $V$ can introduce additional scatter. Here we include a 30% error in linewidth, which is an estimate of the typical combined statistical and systematic uncertainty. Fitting to the primary sample of 9 AGN we find $D = 4.20^{+0.71}_{-0.56}$, strongly supporting our earlier derivation that $T_B \approx M_{BH}/\dot{m}_E$. Adopting slightly different values of the linewidth from other observers changes the fit very little. Thus two apparently quite different observable characteristics of AGN, that is, X-ray variability and optical linewidth, can both be explained as simply depending on $M_{BH}/\dot{m}_E$, thereby linking small scale nuclear accretion properties to larger scale AGN properties.

Our results (see Supplementary Information for further comments) have important implications for understanding the different types of active galaxy as differentiated by their optical linewidths, in particular the NLS1s[28,29]. It is not $M_{BH}$ or $\dot{m}_E$ on their own which define the linewidth, but the ratio $M_{BH}/\dot{m}_E$. The observed small masses of NLS1s are a selection effect as, for an AGN with $M_{BH}$ greater than a few $\times 10^6$ solar masses to produce narrow emission lines, $\dot{m}_E$ must exceed the Eddington limit. Particular orientations, such as face-on, although not ruled out, are not required, nor is an unusual distribution or density of the surrounding gas. Apart from a lower ratio of $M_{BH}/\dot{m}_E$, NLS1s are no different to other AGN.

**Supplementary Information** is linked to the online version of the paper at www.nature.com/nature and is also attached to the end of this document.

**Acknowledgements** This work was supported by the UK Particle Physics and Astronomy Research Council (PPARC). We thank P. Smith for discussions about statistics, P. Arevalo for determining the break timescale in GRS 1915+105 and D. Summons for providing some RXTE lightcurves of Cyg X-1.

**Author Information** Reprints and permissions information is available at www.nature.com/reprints. The authors declare no competing financial interests. Correspondence and requests for materials should be addressed to IM$^c$H (imh@astro.soton.ac.uk)

**Figure 1** Confidence contours for the fit parameters A and B. Here we show the 68% (black), 90% (red) and 95% (green) confidence contours for the dependence of the PSD break timescale on black hole mass and bolometric luminosity. The mass and bolometric indices, *A* and *B*, are defined in the main text. The thick contours refer to the primary sample of 10 AGN with well measured PSD breaks, that is NGC 3227, NGC 3516, NGC 3783, NGC 4051, NGC 4151, NGC 4395, NGC 5506, MCG-6-30-15, Mkn 766 and Akn 564. The best fit values for this fit are $A = 2.17^{+0.3}_{-0.2}$, $B = 0.90^{+0.25}_{-0.2}$ and $C = -2.42^{+0.22}_{-0.25}$ (1$\sigma$ errors), reduced $\chi^2 = 0.85$, 7 degrees of freedom (d.o.f.), for $T_B$ in days, $M_{BH}$ in $10^6$ solar masses and $L_{bol}$ in $10^{44}$ erg s$^{-1}$. For a combined fit to the 10 AGN and the much lower mass Galactic black hole system Cyg X-1, we find $A = 2.10^{+0.23}_{-0.18}$, $B = 0.98^{+0.20}_{-0.16}$ and $C = -2.33 \pm 0.15$ (reduced $\chi^2 = 0.82$, 10 d.o.f.), which is excellent agreement with the AGN on their own. For Cyg X-1 we assume $M_{BH} = 15 \pm 5$ solar masses, but assuming $M_{BH} = 10$, or 20, changes each fit parameter by only 0.025. Including also the Galactic black hole system GRS 1915+105, we find $A = 2.10 \pm 0.15$, $B = 0.98 \pm 0.15$ and $C = -2.32 \pm 0.2$ (reduced $\chi^2 = 0.85$, 11 d.o.f.). The contours for this latter fit are shown as thin lines. Even at the $1\sigma$ (that is, 68% confidence) level, the contours for the AGN on their own, and for the combined AGN and GBH sample, completely overlap (as do the offset constants, $C$).

**Figure 2** Edge-on projection of our sample and the $T_B$-$M_{BH}$-$L_{bol}$ plane. Here we show that the predicted break timescales, $T_{predicted}$, derived by inserting the observed bolometric luminosities and masses into the best fit relationship to the combined sample of AGN and GBHs shown in Fig. 1 (that is, $\log T_B = 2.1\log M_{BH} - 0.98\log L_{bol} - 2.32$), agree very well with the observed break timescales, $T_{observed}$, for all objects. Here we show GRS 1915+105 as a filled maroon star, Cyg X-1 as blue crosses and the 10 AGN as red circles. The low luminosity AGN (LLAGN), NGC 4395, is shown as an open crossed red circle; the other 9 AGN are filled red circles. Although not included in the fit as the upper limit on their break timescales are unbounded, we also plot (filled green squares) NGC 5548, and Fairall 9 and the LLAGN NGC 4258. If the predicted and observed break timescales are identical, then an object will lie exactly on the black line, which is a projection of the best-fit three-dimensional $T_B$–$M_{BH}$–$L_{bol}$ plane and is not a fit simply in two dimensions to the points shown in this figure. We plot the 90% confidence errors on $T_{observed}$ which are usually quoted. In some cases these errors are too small to show beyond the symbols. The only noticeable outlier is NGC 5506 which, on the basis of a narrow Paβ near-infrared line, is classed as an obscured NLS1[30]. However, its current mass estimate, less well determined than for most other AGN, implies $L_{bol}/L_E \approx 2.6\%$, which is surprisingly low for an NLS1. If the mass were a factor of ~5 lower, $L_{bol}/L_E$ would be more normal for an NLS1 and NGC 5506 would sit directly on the plane.

**Figure 3** Correlation of optical emission linewidth with PSD break timescale. We show that the full-width at half-maximum intensity (FWHM) of the Hβ optical emission line in AGN is strongly correlated with the observed PSD break timescale, $T_{observed}$. For $\log T_B = D \log V + E$, we find $D = 4.20^{+0.71}_{-0.56}$, $E = -14.43$ (reduced

$\chi^2 = 1.20$, 7 d.o.f.), with $T_B$ in days and width, $V$, in km s$^{-1}$. The fit is to the 9 AGN with both well measured PSD break timescales k and measured values of $V$. As in Fig. 2, the LLAGN NGC 4395 is shown as an open crossed red circle and the other 8 AGN are filled red circles. NGC 5506 is present in Fig. 2 but missing here, as it is heavily obscured so Hβ is undetectable. As different lines may be produced at different distances from the black hole, alternative lines may not be used. Although not included in the fit, we plot (filled green squares, as in Fig. 2) Fairall 9 and NGC 5548 for whom linewidths are available but whose upper break timescales are unbounded. We note that their present values of $T_B$ are consistent with the fit for the other AGN. Where available we use $V$ as detected in the r.m.s. spectrum[26], as it is a better estimate of the variable component from the BLR and is less contaminated by constant narrow components originating further away. The Hβ broad linewidth of the LLAGN NGC 4395 is poorly determined. If it is removed from the sample, the fit improves slightly ($D = 3.98^{+0.64}_{-0.49}$, $E = -13.62$ with reduced $\chi^2 = 0.96$, 6 d.o.f.). The AGN with logFWHM = 3.72 which lies below the best fit line is NGC 3227. For both NGC 3227 and NGC 4395, there are large time differences between the measurement of $T_B$ and $V$ which may introduce additional scatter. Error bars on $T_B$ are 90% confidence, as in Fig. 2.

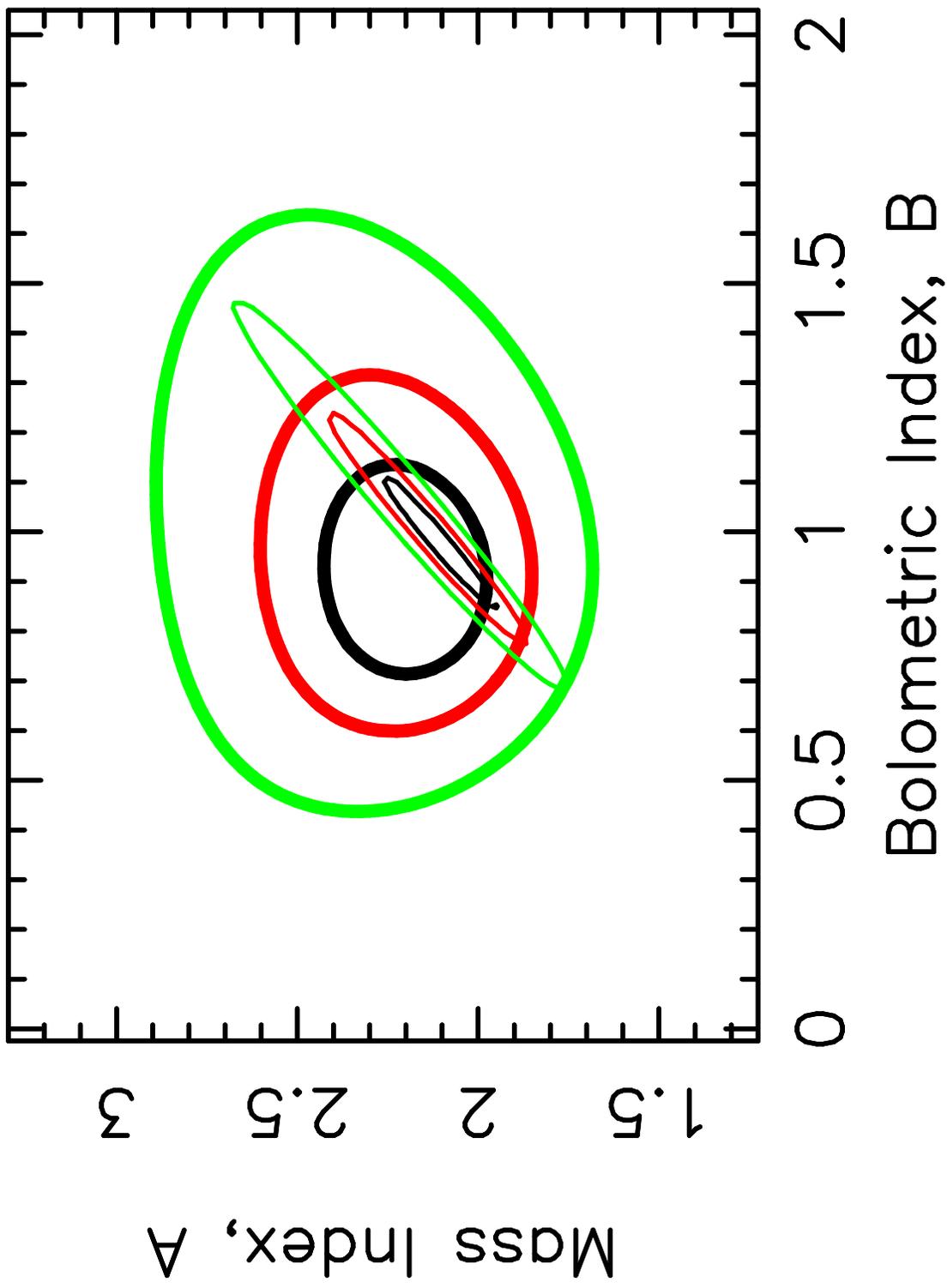

**FIGURE 1.**

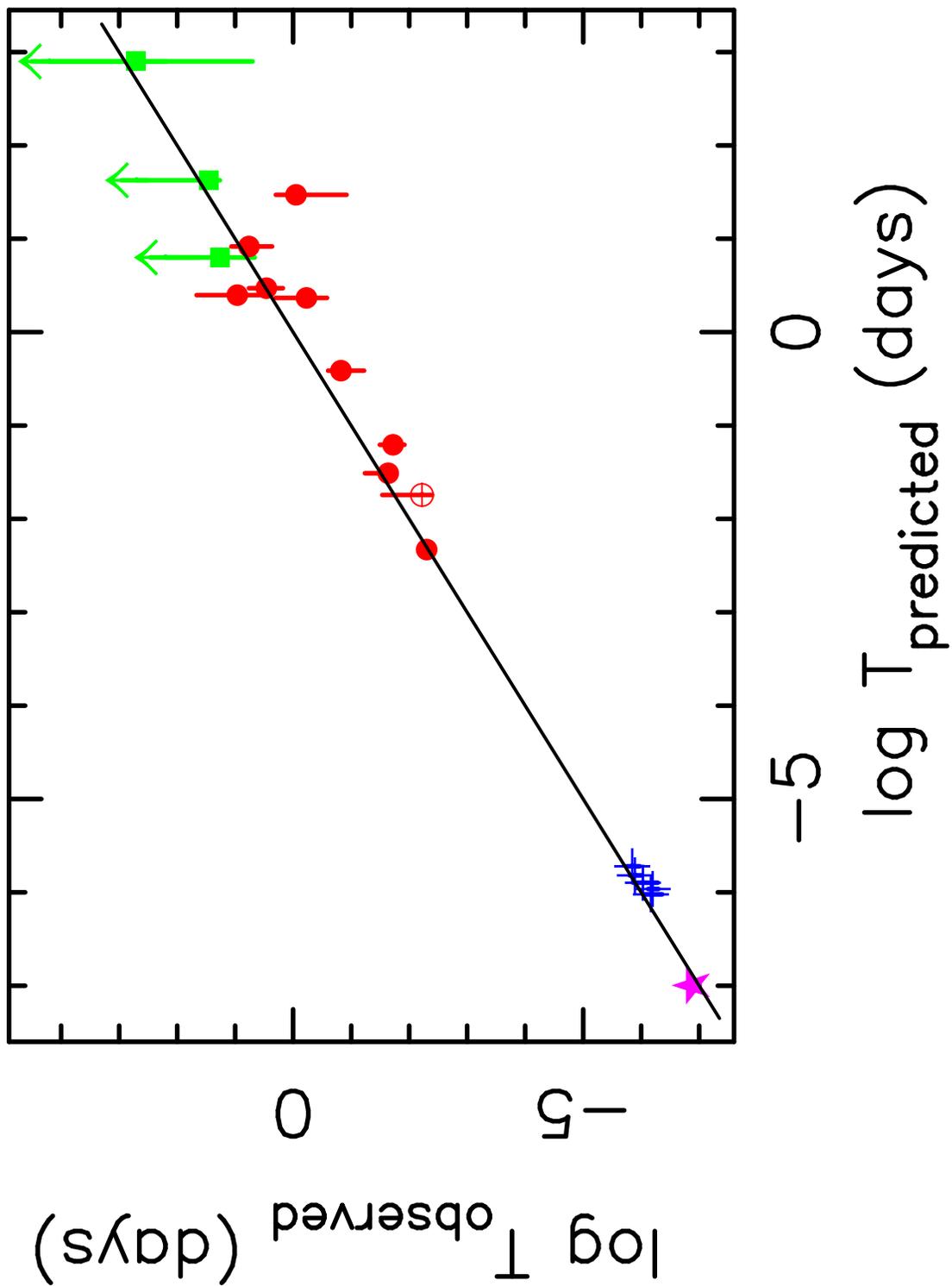

**FIGURE 2**

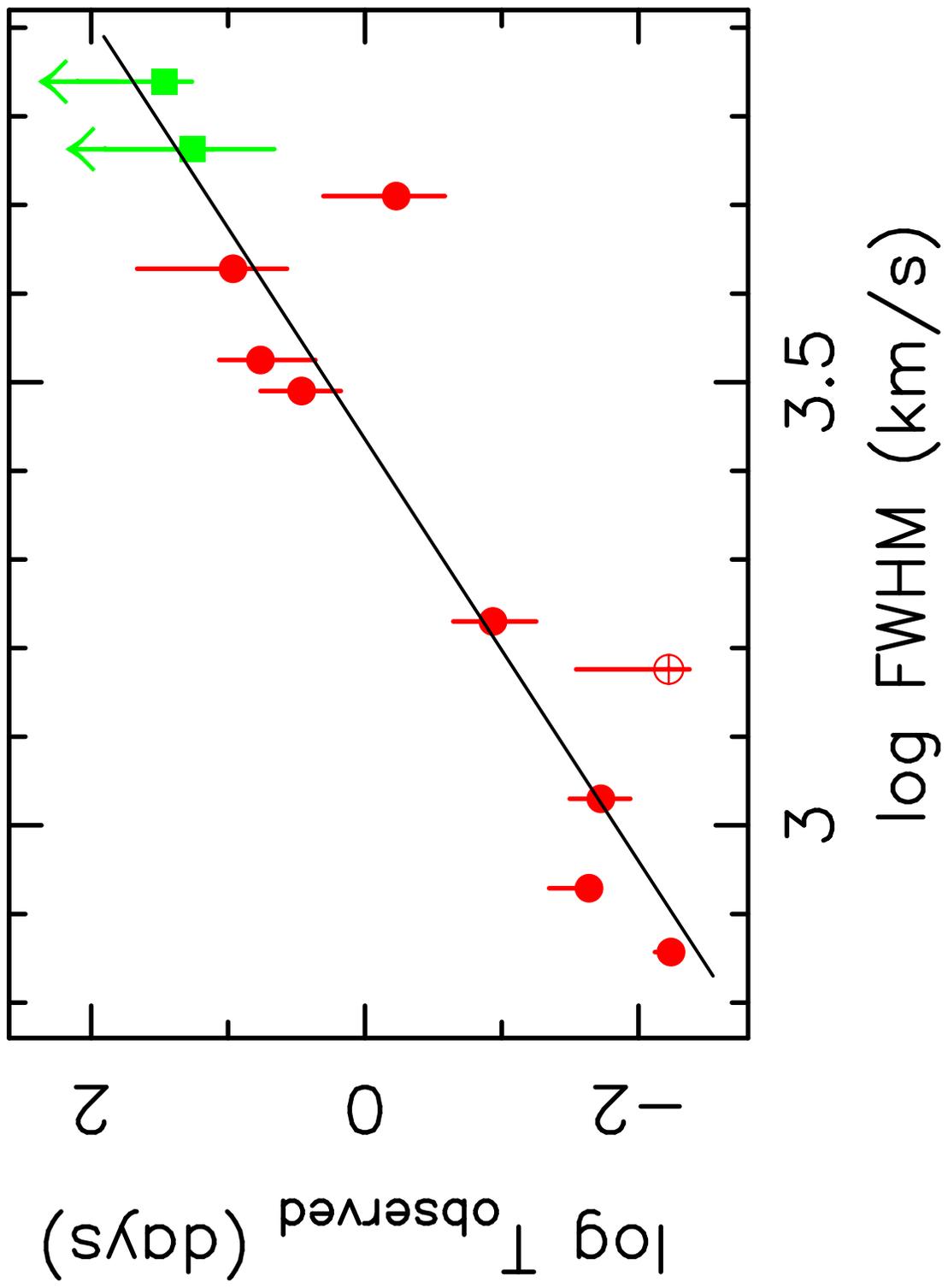

**FIGURE 3**

# SUPPLEMENTARY INFORMATION

## 1. INTRODUCTION

In the above letter we study the scaling relationship between Active Galactic Nuclei (AGN) and Galactic Black Hole X-ray binary systems (GBHs) by means of comparing the way in which their X-ray emission varies[1,2,3,4]. We also show how the widths of the permitted optical emission lines, a fundamental tracer of larger scale AGN properties often used to classify AGN, are intimately related to the way in which we scale timing properties from large to small black holes. In the main paper we show that a characteristic timescale found in both AGN and GBHs scales as the ratio of black hole mass to accretion rate, and the linewidths scale with that same ratio to the fourth power. Thus we link small scale accretion properties to larger scale AGN properties.

Although possible similarities between AGN and GBHs have been muted ever since the late 70's or early 80's when it was first realised that they were both black hole systems[5], comparison of their X-ray variability properties provided the first quantitative method for this comparison. More recently considerable attention has been devoted to the jet properties of black hole systems, eg their radio luminosities, and a strong scaling has been shown by means of comparing radio and X-ray luminosities[6,7]. However here we concentrate on the considerable insight which can be gained regarding the scaling of the between AGN and GBHs by studying their X-ray timing similarities.

In this supplementary information we fill in some technical details which are not presented in the main paper but do not reproduce the main paper.

## 2. PSD PARAMETERISATION

In order to compare timing properties, we must somehow quantify the timing observations in such a way that we can measure similar parameters from observations of different sources, and so compare the sources. There are a number of ways in which timing can be quantified but the starting data are the same in all cases. These data consist of measurements of intensity, or flux, as a function of time and are commonly referred to as `lightcurves'. Although lightcurves can be quantified in the `time domain', eg by calculating `structure functions' or `fractal plots'[8], the most commonly used method is to Fourier analyse the lightcurves to determine their powerspectral densities (PSDs)[9,10,11], ie the power of the variability, $P(\nu)$, on frequencies, $\nu$, or timescales, $T (= 1/\nu)$. The power is the square of the amplitude of typical sinusoidal variations of frequency $\nu$ within the lightcurves. Historically, AGN researchers have referred to `break timescales' whereas GBH researchers prefer to use frequencies. We apologise for any possible confusion.

Galactic black hole binary systems, being much closer than AGN, are much brighter in X-rays and so it is possible to measure their lightcurves to high precision even on very short (millisecond) timescales. Their powerspectra are therefore much better determined than those of AGN and so are involved in the classification of the different `states' of GBHs.

The classification of GBHs is not unambiguous and different researchers use different terms. Powerspectral properties seem to correlate more with X-ray spectral hardness than with intensity and so the terminology `hard' and `soft' is now more common than `high (intensity)' or `low'. In the hard state the large majority of the X-ray flux comes from a highly variable component and so the hard state powerspectrum is easy to measure. In the soft state that highly variable component still exists, although with different variability characteristics, but is heavily diluted by a more luminous quiescent component. Thus, in the soft state, fractional variability is often very low and it is hard to quantify soft state PSDs. Intermediate states, of course, also exist. The intensity of intermediate states can be very high, when they are sometimes also referred to as `very high states'.

Where measureable, eg in Cygnus X-1, soft state pds have a powerlaw form on long timescales, ($P(\nu) \propto \nu^{-\alpha}$ with $\alpha \approx 1$) which steepens, or breaks, at a particular timescale which is referred to as the `break timescale', $T_B$, to a powerlaw of slope $\alpha \approx 2$, or steeper. There are no further breaks on longer timescales. The break is sometimes parameterised as an exponential cut-off[12] and, in Galactic systems where the break is at high frequencies and frequency coverage above the break is often limited, is often difficult to tell the difference between the broken powerlaw and the exponentially cut-off parameterisation[13].

The PSD in the hard state is more complex, being best described by the sum of two or more components with Lorentzian shapes, but it can also be approximated by a doubly-broken powerlaw[9,10,14]. In this parameterisation we again see the break, at high frequencies, from powerlaw slopes of 2 (or steeper) to 1 but, in addition, there is a second break, at longer timescales, between slopes of 1 and 0. The slope of $\alpha \approx 1$ only extends for about a factor of 30 in time (or frequency), ie about 1.5 decades, and so provides a method of distinguishing between soft and hard states. For some AGN this method has been used to determine that they are soft states but for the majority, frequency coverage is not adequate to distinguish soft from hard states. However all AGN display the break from slopes of 1 to 2.

In GBHs the transition from hard to soft state[15] occurs at ~2% of the maximum possible, or Eddington, accretion rate (i.e. $\dot{m}/\dot{m}_E$ ~0.02). However if that rate depends on how close towards the black hole the optically thick disc can reach before evaporating, it may be lower in higher mass black holes (AGN) with cooler discs. Most of the present AGN sample have $\dot{m} > 0.02 \dot{m}_E$ and so are presumably soft state objects although NGC4395 and NGC4258, which have very low accretion rates ($\dot{m}/\dot{m}_E$ ~ 0.001) are more likely to be hard state. It is possible that Akn564[4,14] may be a very soft state object but further work (M$^c$Hardy et al in preparation) is required to clarify the shape of its PSD. However it also shows the high frequency break from slopes of approximately 2 to 1. In this paper our conclusions are therefore based on the timescale associated with the break from slope 2 to 1, independent of likely state, which can be seen in all AGN and GBH PSDs.

In the Lorenztian description of the hard, or very high, state PSD, the components are centred approximately at the powerlaw break timescales, and their central timescales vary with flux in a similar, though not absolutely identical, manner[14]. There are good indications that the central timescales of some Lorentzian components may vary with accretion rate in a manner consistent with that derived from the break from slope 2 to 1 in this paper[16]. However that conclusion is based on radio observations as well as X-ray observations and requires some assumptions based on theoretical modes for jets in order to estimate the accretion rate. More work is required to determine whether timescales and accretion rates derived in this manner can be amalgamated with the simple PSD break timescales and accretion rates derived simply from bolometric luminosities which we use here.

## 3. FITTING PROCEDURE

To determine the best-fit values of A,B and C, we define a standard $\chi^2$ figure of merit in the same way as the `nukers'[17] and perform a simple parameter grid search to determine the minimum value of $\chi^2$. As 90% confidence errors are usually quoted on $T_B$, we have examined the confidence contours (published and unpublished) for $T_B$ and find that the errors are roughly Gaussian and so have renormalised to $1\sigma$ errors for use in the $\chi^2$ fitting. As the errors on $T_B$ are also usually asymmetric we take the error, for a given set of trial parameters, which lies in the direction from the observed to the predicted value of $T_B$. Errors on $L_{bol}$ are not usually provided so we assume a typical ~50% error which corresponds to the typical spread in estimated $L_{bol}$ for the same AGN from different observers. We have implemented other algorithms for dealing with asymmetric errors[18,19] but find that the results do not depend significantly on the method used.

For $T_B$ we take the timescale associated with the break, in the powerlaw approximation to the PSD shape, between slopes of 1 and 2 which is found in both AGN and GBHs. The PSDs of AGN are not well enough defined to allow the more complex parameterisation in terms of Lorentzians which can be applied to GBHs.

## 4. DATA RELEVANT TO THE $T_B$-M-$L_{BOL}$ FITTING

### 4.1 AGN

We mostly take the data for $T_B$, $M_{BH}$ and $L_{bol}$, as tabulated by Uttley and M$^c$Hardy[20] and refer readers to that paper for the full data and for the relevant references. Most masses are derived from reverberation mapping, mainly by Peterson et al[21] with bolometric luminosities from Wu and Urry[22]. However the mass for NGC5506 ($10^8$ solar masses) is derived, less accurately, from the width of the [OIII] lines and so we assume a factor 2 mass error, typical of that method[23]. We note that, on the basis of its Pa near-IR line (FWHM ~ 1800 km s$^{-1}$) NGC5506 is classed as an obscured NLS1[24]. However its current mass estimate implies $L_{Bol}/L_E$~2.6%, which is surprisingly low for an NLS1. If the mass were a factor ~5 lower, its $L_{Bol}/L_E$ would be similar to those of other NLS1s and it would sit directly on the mass/accretion rate/break-timescale plane.

For Mkn766[25] we adopt a similar error based on the width of the stellar absorption lines. For NGC3227 we take the recent mass based on detailed stellar dynamics[26] which should be more accurate than that based on reverberation mapping[21]. For NGC4395 we take the mass as listed by Uttley and M$^c$Hardy[20] which is based on the very low stellar velocity dispersion ($\tilde{\sigma} < 30$ km s$^{-1}$) and very low nuclear optical emission[27,28] rather than the x10 higher estimate, based on short HST reverberation mapping observations[29]. We suspect that the former is more reliable as only one cycle of variation is covered in each HST visit so the line and continuum variations, upon whose relative lag the mass is based, may not be physically related.(E.g see the discussion regarding the likelihood of chance association of X-ray and optical variations in NGC4051 as observed by XMM-Newton[30]).

**4.2 Cygnus X-1**

An extensive compilation of powerspectra are provided for Cyg X-1 in the soft and transitional states[12]. These PSDs have a typical time resolution of ~1ksec. A variety of models are fitted:

1 = only two Lorentzians are present,

2 = two Lorentzians and cut-off power-law,

3 = refitted results (two Lorentzians and cut-off power-law present),

4 = first Lorentzian and cut-off power-law only,

5 = purely cut-off power-law.

Model 5 is almost exactly the same model which we use to fit AGN PSDs and so these Cyg X-1 data are directly comparable with the AGN data. In models 2, 3 and 4, a cut-off powerlaw is fitted to the data but the Lorentzian components dominate the PSD. The central frequency of the higher frequency Lorentzian is not far different from that of the cut-off frequency. Thus the cut-off frequency is not so cleanly measured. The model 5 fits come almost entirely from the period late 2001 to late 2002 when the 2-10 keV X-ray flux was high and the radio flux low, so the spectral fits to the same data[31] do provide a good estimate of the total bolometric luminosity. We have measured the average value of the powerlaw slope in model 5 and find -1.00±0.068, the same value as is found in AGN. As the measured cut-off frequency can be affected by incorrect slope measurements we therefore reject all data with slopes more than 1 sigma from the mean.

In order to obtain the best estimate of the bolometric luminosity at the time of the PSD measurements, we take the published high quality spectral fits based on the eqpair model[31]. As the temporal periods sampled by the spectral fits are not exactly the same as those of the PSD measurements, we take only data where the two time periods overlap. We bin the resulting data into 5 luminosity bins.

The results of fitting are described in the main paper, and the Cyg X-1 data are plotted in Fig 2. However we note here that in the fit which includes the AGN and Cyg X-1, the mass is the same for each measurement of bolometric luminosity and break timescale for Cyg X-1, ie we are in general down to two rather than three variables. We therefore assign a reduced weight (2/3), and a similarly reduced contribution to the degrees of freedom, to the Cyg X-1 data points, compared to unity weight for the AGN and for GRS1915+105. We note that, if we take only one of the Cyg X-1 data points and assign unity weight, and fit that together with the AGN and GRS1915+105, we obtain almost identical fit parameters and a very similar quality of fit.

We assume a mass of 15 solar masses. This mass is intermediate between the most recent mass determination[32] of 20 solar masses and the previous determination of 10 solar masses[33]. It isn't clear which estimate is most accurate but the exact choice makes little difference to our fit.

**4.3 GRS1915+105**

GRS1915+105 is a Galactic black hole X-ray binary system which is persistently in a very luminous and high accretion rate condition. Sometimes it has strong radio emission and on other occasions it is radio quiet and so its broad band X-ray flux provides an accurate measurement of its accretion rate. A study of its spectrum and variability in the radio quiet phase, referred to as a `type 1 state', has recently been published[34]. In this state (see Figs.2a, 2b and 2c of reference 33), the powerspectrum at high frequencies (>10 Hz) is dominated by broad-band noise, with a high frequency cut-off, very similar to that seen in the soft state PSDs of Cyg X-1, as described above. In this case, although a Lorentzian component is also present, the cut-off in the broad band noise is at a much higher frequency (~100Hz) than the Lorentzian and so it is possible to measure, quite accurately, the cut-off frequency and the PSD powerlaw slope below the cut-off.

The published paper[34] approximates the cut-off with a zero-centred Gaussian component. For consistency with our other data we have therefore extracted the 6 type 1 state data from the RXTE archive and refitted the 2-13 keV PSDs above 20 Hz with model 5 described above. These data cover only a small flux range. We find that they are well described by a powerlaw of index $-1.06^{+0.06}_{-0.06}$ and an average cut-off frequency of $92^{+17}_{-17}$ Hz. The value of the powerlaw index is in agreement with the mean value listed above for model 5 of Cyg X-1. (If we fix the slope at -1.00, the cut off frequency barely changes, to 85±15Hz). Neither the 3-150 keV X-ray flux, or the cut-off frequency, change significantly from observation to observation and so here we take the average values (92.2 Hz and 1.98±0.28 x $10^{-8}$ ergs cm$^{-2}$ s$^{-1}$). In this state, GRS1915+105 has a rather hard spectrum and so the 3-150 keV flux contains the large majority of the bolometric flux.

We convert the average flux to a bolometric luminosity assuming a distance of 11kpc. There has been a question as to the true distance to GRS1915+105, with a closer distance of 6.5kpc being proposed[35].. However the majority opinion remains that the true distance is approximately 11kpc[36] and we follow the majority opinion. We similarly accept the majority view on the mass of the black hole of 14±4 solar masses[36]. The resultant values are included in our fitting process. The mean value of the flux used here corresponds to an accretion rate of 16% $\dot{m}_E$, which is significantly higher than the range of accretion rates probed by Cyg X-1 (1-3%) and so provides a very useful test of the quality of the fit.

**4.4. GX339-4 and XTEJ1550-564**

Here, and in the main paper we note that for GBHs in the hard state where the PSD is very well defined, Lorentzians provide a better description of the PSD than breaking powerlaws. Thus, although not yet definitively proven, the timescales associated with Lorentzian components may be a better tracer of accretion rate variations than powerspectral break timescales. However, in combination with measurements of the accretion rate based on radio measurements, together with jet modelling, it has already been shown that in GX339-4 in the hard state, the timescale associated with the highest frequency Lorentzian, $\nu_{high}$, varies approximately as the inverse of accretion rate, as we find here in the main paper.

From data kindly provided by Done and Gierlinski[14] for the GBH XTEJ1550-564 we have calculated that, in 2002 when the source was in a steady hard state, $\nu_{high} \sim S_X^{0.45 \pm 0.05}$, where $S_X$ is the medium energy X-ray flux. [Done and Gierlinski refer to $L_{bol}$ rather than $S_X$ but derive $L_{bol}$ by simple linear transformation from X-ray flux.] In hard state, jet dominated, systems[37], $S_X \sim \dot{m}^2$. Thus for the 2002, hard state, $\nu_{high} \sim \dot{m}^{0.9 \pm 0.1}$, giving the same dependence of timescale on $\dot{m}$ as for the samples discussed in the main paper.

**5. FITTING OF LINEWIDTH AGAINST PSD BREAK TIMESCALE**

We take the FWHM of Hβ as detected in the rms spectrum by Peterson et al[21] for Fairall 9, NGC3227, NGC3516, NGC3783, NGC4051, NGC4151 and NGC5548; for NGC5548 we take an average value of 5800 km s$^{-1}$. However for Akn564, Mkn766, MCG-6-30-15 and the LLAGN NGC4395, only single epoch linewidths are available[38,39,40]. The Hβ emission line in NGC4395 is dominated by a strong narrow component and the width of the very weak broad component is therefore particularly poorly determined[40] (~1500 km s$^{-1}$). We note that a much narrower width (FWHM 442 km s$^{-1}$ ) is listed for the total Hα line[41]. It is therefore not clear whether the deviation of NGC4395 from the strong linewidth/break timescale correlation is a result of measurement error, or indicates a true physical difference between the BLRs of LLAGN and more luminous AGN. Neither NGC5506 nor the LLAGN NGC4258[42] are plotted as Hβ is obscured . Interestingly we note that if Hβ (obscured) in NGC5506 has the same width as Paβ (visible), NGC5506 would lie on the best-fit line.

For both NGC3227 and NGC4395, there are large time differences between the measurement of $T_B$ and $V^{20,21,40,43}$ which may very well contribute can introduce additional scatter into the relationship.

## 6. FURTHER IMPLICATIONS OF OUR RESULTS

### 6.1 Extrapolation of the break timescale/linewidth relationship to Galactic Systems

We can extrapolate the break timescale/linewidth relationship to GBHs. A typical break frequency of 15Hz (8x10$^{-7}$ days) for Cyg X-1 in the soft state translates to a velocity of 95 km/s, and a break frequency of 1Hz (9x10$^{-6}$ days) for the low state translates to 180 km/s. These values are factors of 2 or 3 lower than are typically found in GBHs[44,45] but we should note that the error in the extrapolation is of a similar value. Further work is required to determine whether the emission line regions in AGN and GBHs share a common origin.

## 6.2 Spin

If our assumption of a typical 50% error on $L_{bol}$ is correct, then the quality of our fit between break timescale, mass and $L_{bol}$ implies that no other source of error, such as is commonly invoked in fits of this type to achieve a reduced $\chi^2$ close to unity, is required. Thus spin cannot have as large an affect on break timescale as accretion rate. Spin ultimately limits the last stable orbit that matter can inhabit close to the black hole, ie the smallest possible inner edge of the accretion disc. Therefore if the break timescale is associated with the inner radius of the accretion disc then either the X-ray emitting region is very far out in the accretion disc, which is not generally considered likely, or else most X-ray luminous black holes are spinning quite rapidly. In the latter case the last stable orbit would be small and so variations in radius caused by variations in accretion rate would not hit the end-stops of the last stable orbit. Rapid spin is consistent with the broad iron lines seen in some AGN[46].

There are a number of assumptions in this derivation, and so it should not be considered as firm, but if future observations improve our confidence in our error estimates, the above timing argument may be useful in establishing the spin of X-ray bright black holes.

## 6.3 Mass, Linewidth and Accretion Rate

In the main letter we derive two empirical equations:

$$\log T_B = 2.1 \log M_{BH} - 0.98 \log L_{Bol} - 2.32$$

$$\log T_B = 4.2 \log V - 14.43$$

Remembering that $M_{BH}$ is in units of $10^6$ solar masses and $L_{Bol}$ is in units of $10^{44}$ ergs s$^{-1}$ and V is in km s$^{-1}$, we can combine these equations and write:

$$\log M_{BH} = 0.875 \log \dot{m}_E + 3.75 \log V - 10.71$$

Consider, then, a narrow line Seyfert galaxy with V=1000 km s$^{-1}$. Unless $\dot{m}_E >$ 1, then the black hole mass must be less than 3.5 x 10$^6$ solar masses. We caution that there are considerable uncertainties in all coefficients, so this latter equation should be treated cautiously, however it does give us some idea of the mass range allowed for AGN of particular linewidths and accretion rates.

Acknowlegements

We thank Chris Done and Marek Gierlinski for providing us with the detailed numbers for XTEJ1550-564 .